\documentclass{article}

\usepackage{arxiv}

\usepackage[utf8]{inputenc} 
\usepackage[T1]{fontenc}    
\usepackage{hyperref}       
\usepackage{url}            
\usepackage{booktabs}       
\usepackage{amsfonts}       
\usepackage{nicefrac}       
\usepackage{microtype}      
\usepackage{lipsum}
\usepackage{graphicx}
\graphicspath{ {./images/} }

\usepackage{cite}
\usepackage{amsmath,amssymb,amsfonts,bm}
\usepackage{algorithmic}
\usepackage{graphicx}
\usepackage{textcomp}
\usepackage[capitalise]{cleveref}
\usepackage{float}
\usepackage{tikz}

\usetikzlibrary{shadows}
\usetikzlibrary{shadows.blur}
\usetikzlibrary{arrows,shapes,backgrounds,patterns,fadings,decorations,positioning}

\newtheorem{remark}{\bf Remark}
\newtheorem{assumption}{\bf  Assumption}
\newtheorem{definition}{\bf  Definition}
\newtheorem{lemma}{\bf  Lemma}
\newtheorem{theorem}{\bf  Theorem}

\title{Distributed Learning Consensus Control for Unknown Nonlinear Multi-Agent
Systems based on Gaussian Processes\thanks{A preprint submitted to IEEE CDC2021.}}

\author{
 Zewen Yang \\
 Department of Electrical and Computer Engineering\\
 Technical University of Munich\\
 Germany, Munich 80333   \\
 College of Intelligent Systems Science and Engineering\\
 Harbin Engineering University\\
China, Harbin 150001 \\
  \texttt{yangzewen@hrbeu.edu.cn} \\
   \And
 Stefan Sosnowski \\
 Department of Electrical and Computer Engineering\\
 Technical University of Munich\\
 Germany, Munich 80333   \\
  \texttt{sosnowski@tum.de} \\
     \And
 Qingchen Liu \\
 Department of Electrical and Computer Engineering\\
 Technical University of Munich\\
 Germany, Munich 80333   \\
  \texttt{qingchen.liu@tum.de} \\
     \And
 Junjie Jiao \\
 Department of Electrical and Computer Engineering\\
 Technical University of Munich\\
 Germany, Munich 80333   \\
  \texttt{junjie.jiao@tum.de} \\
     \And
 Armin Lederer \\
 Department of Electrical and Computer Engineering\\
 Technical University of Munich\\
 Germany, Munich 80333   \\
  \texttt{armin.lederer@tum.de} \\
     \And
 Sandra Hirche \\
 Department of Electrical and Computer Engineering\\
 Technical University of Munich\\
 Germany, Munich 80333   \\
  \texttt{hirche@tum.de} \\
}

\begin{document}
\maketitle

\begin{abstract}
 In this paper, a distributed learning leader-follower consensus protocol based on Gaussian process regression for a class of nonlinear multi-agent systems with unknown dynamics is designed. We propose a distributed learning approach to predict the residual dynamics for each agent. The stability of the consensus protocol using the data-driven model of the dynamics is shown via Lyapunov analysis. The followers ultimately synchronize to the leader with guaranteed error bounds by applying the proposed control law with a high probability. The effectiveness and the applicability of the developed protocol are demonstrated by simulation examples.
\end{abstract}


\section{Introduction}

Efficient control laws for nonlinear systems typically require precise knowledge of the system dynamics. This knowledge is especially important for model-based control techniques such as feedback linearization, backstepping or model predictive control \cite{oriolo2002wmr,van2018adaptive,shen2017trajectory}. Additionally, unknown interference also influences the stability of the control systems. Therefore, data-driven control approaches that identify the unknown system model according to generated data have received a lot of interest recently. In this paper we specifically consider Gaussian process regression (GPR), a data-driven learning approach. It is a tool widely used for data-based control, especially in learning and modeling complex nonlinear systems \cite{umlauft2017learning}. GPR provides several advantages: It provides uniform error bounds, which give the guarantees of safe control based on the data-driven model \cite{lederer2019uniform}. In comparison to neural networks (NN), GPR provides better results for small training data sets. Another advantage is the flexibility coming with its nonparametric nature. Only minimal prior knowledge is required \cite{umlauft2019feedback}. Yet, still most studies on data-driven control based on GPR are committed to single systems.

Recently, the control of multi-agent systems (MAS) has received much attention, see e.g. \cite{cao2012overview,oh2015survey,abdulghafor2018overview,qian2018output,li2019multilayer}. Typically, in these works the dynamics of MAS is assumed to be known. However, it is often challenging to obtain the accurate dynamics and environmental disturbances that are added to the uncertainties of the system model. It is therefore of practical significance, to resolve the  control problem of MAS with unknown dynamics. A few works employ NNs to approximate the unknown nonlinear function in MAS, see e.g. \cite{chen2014adaptive,el2014neuro,zou2020finite}. A drawback of this approach is that there is no guaranteed bound of the approximation error. In addition, compared with the nonparametric nature of GPR, resolving the optimization problem of the weight parameters in NN is time consuming. 

A related field of research to our problem setting is the area of distributed learning algorithms, however, without considering control. A GP decentralized data fusion algorithm is proposed in \cite{ouyang2018gaussian}. The design and evaluation of a distributed method for exact GP inference presented in \cite{nguyen2019exact} achieves true model parallelism using simple, high-level distributed computing frameworks. Under the collective online learning of GP framework \cite{hoang2019collective}, the online learning algorithms can fuse and update online GP models efficiently with varying correlation structures. The communication-aware GP algorithm allows a network of robots to collaboratively learn about the unknown functions with each other \cite{yuan2020communication}. Very recently, a distributed model predictive control approach for MAS with GPR has been proposed \cite{le2020gaussian}. However, the work mainly focuses on the cooperative optimization problem. To the best of our knowledge, there exists no consensus control approach with GPR that guarantees the convergence of tracking errors in unknown nonlinear MAS. 

The main contribution of this paper is the design of a learning-based leader-follower consensus protocol for unknown nonlinear MAS based on GPR. We propose a distributed non-parametric learning approach to model the unknown residual nonlinear dynamics. Instead of learning individually in each follower, we design a novel distributed learning approach based on GPR for the MAS, where the followers share the information of prediction and then aggregate the individual predictions of their neighbors. Using a suitable Lyapunov function, we prove that the multi-agent system is stable and the proposed control law guarantees the followers ultimately synchronize to the leader in a guaranteed errors bound with a high probability.

The remainder of this article is structured as follows: Preliminaries including the relevant notation and basic graph theory are stated in Section \uppercase\expandafter{\romannumeral2} followed by the problem formulation in Section \uppercase\expandafter{\romannumeral3}. In Section \uppercase\expandafter{\romannumeral4} the proposed consensus tracking control protocol employing a distributed learning approach based on GPR is presented, and stability for the resulting closed-loop MAS is proven. The numerical simulation demonstrates the effectiveness of the proposed approach in Section \uppercase\expandafter{\romannumeral5} followed by a conclusion in Section \uppercase\expandafter{\romannumeral6}.

\section{Preliminaries}

\subsection{Notation}

 In this paper, we let $\mathbb{R}^{m}$ denote the $m$-dimensional Euclidean space and $\mathbb{R}^{m\times n}$ denote the set of $m\times n$ real matrices, respectively. The set of real number is denoted by $\mathbb{R}$,  $\mathbb{R}_{0,+}/\mathbb{R}_+$ indicates the set of real positive numbers with/without zero. The transpose of a vector or matrix $A$ is given by $A^\top$. The smallest/largest eigenvalues of a matrix $A$ are denoted as $\lambda_{\min}(A)$ and $\lambda_{\max}(A)$, respectively. The $n\times n$ identity matrix is $\emph{I}_{n}$. The symbol $\otimes$ denotes the Kronecker product. The Euclidean norm of a vector, and the matrix norm induced by the Euclidean norm, are denoted by $\|\cdot\|$. $\mathcal{N}(\mu,\sigma^2 )$ denotes a Gaussian distribution with mean $\mu$ and variance $\sigma^2$.

\subsection{Graph Theory}
 We model the interactions among the leader and $n$ followers by an undirected graph, where the leader is denoted by node $0$ and the followers are denoted by nodes $ 1,\dots, n $. We use $G =({V},{E},{A})$ to describe the interactions among the $n$ follower agents with node set $V=\{v_1,\cdots,v_n\}$, edge set $E\subseteq V\times V$ and adjacency matrix $A=\{a_{ij}\}$. An ordered edge set of $G$ is $e_{ij}=(v_i,v_j)$. The adjacency matrix $A=\{a_{ij}\}$ is the $n\times n$ matrix given by $a_{ij}=1$, if $e_{ij}\in {E}$ and $a_{ij}=0$, otherwise. We assume that $a_{ii} = 0$, i.e. graph $G$ does not contain self-loops. For an undirected graph, $e_{ij}\in E \Leftrightarrow e_{ji}\in E$ and $a_{ij} = a_{ji}$. Diagonal matrix $ D = \mathrm{diag}\left \{ d_{11},d_{22},\dots d_{nn} \right \} $ is the degree matrix of $G$, where the element of $ d_{ii} = \sum_{j=1}^{n} a_{ij}$. The Laplacian matrix of the graph $G$ is defined as $ L= D-A $. The graph for all the leader-follower agents is denoted as $\bar{G}=(\bar{V},\bar{E},\bar{A})$ with the node set $ \bar{V} = V \cup \left \{ 0 \right \}$. We use a diagonal matrix $B = \mathrm{diag}\left \{ b_{11},b_{22},\dots b_{nn} \right \} $ to describe the connection between the $i$-th agent and the leader. The entry $b_{ii} =1$ if the $i$-th agent connects to the leader and $b_{ii} = 0$ otherwise.

\begin{lemma}[ \cite{hong2006tracking}] \label{lem_tidleL}
   Let $L = \left ( l_{ij} \right ) \in \mathbb{R}^{n\times n} $ be a Laplacian matrix of a connected undirected graph. Then the matrix
   \[\tilde{L} = \begin{bmatrix}
 l_{11}+b_{11} & \cdots  & l_{1n}\\
 \vdots  & \ddots  & \vdots \\
 l_{n1} & \cdots  & l_{nn}+b_{nn}
\end{bmatrix}  
\]
is positive definite, if there exists an $i$ such that $b_{ii}>0$.
\end{lemma}

\section{Problem Formulation}

In this paper, we consider a  nonlinear MAS, which consists of $n$ homogeneous followers and one virtual leader. The dynamics of the $i$-th follower is given by 
\begin{equation} \label{followers} 
\dot{x}_i  = f\left( {x_i} \right) + u_i + h \left ( {{x_i}}   \right ), \quad i= 1,2,\dots, n, 
\end{equation}
where $ {x_i} = \left [ x_{i1},\dots, x_{im} \right ]^\top \in \mathbb{X}$ in a compact set $\mathbb{X} \subset  \mathbb{R}^m  $ represents the state vector of the $i$-th follower, $u_i$ represents the input of the $i$-th follower, $f\left( {x_i} \right): \mathbb{X} \to \mathbb{R} ^m $ represents the {\em unknown} dynamics of the $i$-th follower, $h \left ( x_i \right ): \mathbb{X} \to \mathbb{R} ^m  $ is an {\em unknown} disturbance.

\begin{remark} \label{rem_residual_dynamics}
The unknown functions $f(\cdot )$ and $h(\cdot )$ are not merged together on purpose as it allows to extend the proposed approach straightforwardly to the case of \emph{partially} unknown dynamics. 
\end{remark}

\begin{assumption} \label{bounded_fi}
The unknown nonlinear functions $f \left( \cdot  \right) $ and $h \left ( \cdot  \right )$ are globally bounded and continuously differentiable.
\end{assumption} 

Differentiability is a natural assumption for many physical systems. Additionally, bounded functions $f \left( \cdot \right) $ and $h \left ( \cdot \right )$ would be automatically guaranteed (due to the differentiability) if the set $\mathbb{X}$ was bounded\cite{umlauft2019feedback}.

The dynamics of the virtual leader is given by
\begin{equation}
    \label{leader}  \dot{{x}}_l = f_l\left( {{x}}_l ,t \right),
\end{equation}
where $ {{x}}_l \in  \mathbb{X} $ represents the state vector of the virtual leader, and $f_l\left( {{x}}_l   ,t \right): \mathbb{X} \to \mathbb{R} ^m $ represents the {\em known} dynamics of the leader.
\begin{assumption} \label{bounded_fl}
The nonlinear function $f_l \left( {x}_l, t \right) $ of the leader dynamic is a bounded and continuous function. There exists a positive constant $\Bar{f}_l $ satisfying $ \left \| f_l\left ( {\bm{x}}_l, t  \right )  \right \| < \bar{f}_l $, for all $t$.
\end{assumption}

The nonlinear function $f_l \left( {x}_l, t \right) $ represents the dynamics of the leader (to be tracked by the followers), it is therefore reasonable to assume it to be continuous and bounded.

We define the tracking error between the $i$-th follower and the leader to be 
\begin{equation}\label{tracking_error}
    e_i =x_i - x_l.
\end{equation}
From \eqref{followers} and \eqref{leader}, the error dynamics is obtained as follows
\begin{equation} \label{e}
    \dot{e}_i = f \left ( x_i\right ) +h\left ( x_i \right ) - f_l\left ( x_l,t \right ) + u_i .
\end{equation}
Similarly, for each agent, we define the consensus error to be
\begin{equation}
   \begin{split}
       \xi_i   &= \sum_{j=1}^{n} a_{ij}\left ( x_i   - x_j    \right ) + b_{ii} \left ( x_i   - x_l    \right ) \\
       &= \sum_{j=1}^{n} a_{ij}\left ( e_i - e_j \right ) +  b_{ii} e_i  ,
   \end{split} 
\end{equation}
where $a_{ij}$ is the $ij$-th entry of the  adjacency matrix $A_F$ of the communciation graph among the followers, and $b_{ii}$ is the $ii$-th  entry of the diagonal matrix $B$. The equality $b_{ii} = 1$ means that the leader shares its state information with the $i$-th follower.

The MAS  \eqref{followers} and \eqref{leader} is to be interconnected by a distributed protocol of the form
\begin{equation} \label{controller}
u_i   = -k_i \xi_i - \tilde{\mu}_i\left ( x_i \right )  ,\quad i= 1, \dots, n,
\end{equation} 
where $k_i$ are control gains to be designed, and $\tilde{\mu}_i$ are predictions of the unknown dynamics ${\tau}\left ( x_i \right )  = f\left ( x_i \right ) +  h \left( x_i \right)$, to be determined later.

\begin{assumption} \label{ass_graph}
The communication graph $G$ among the followers is assumed to be a connected undirected graph, and the leader is assumed to share its state information with at least one of the followers. 
\end{assumption}

\begin{definition}[Practical consensus]
The consensus protocol \eqref{controller} is said to achieve practical consensus for the leader-follower MAS \eqref{followers} and \eqref{leader} if the states of the leader and the followers satisfy:
\begin{equation*}
    \lim_{t\rightarrow\infty}\|x_i(t) - x_l(t)\|\leq\delta,~~i=1,\ldots,n,
\end{equation*}
where $\delta$ is a small but positive constant.
\end{definition}

\begin{remark}
\textit{Practical consensus} indicates that the leader-follower consensus error lies within an interval, i.e. zero.
\end{remark}

The objective of this paper is to design distributed protocols of the form \eqref{controller}  such that practical consensus is achieved, i.e., the tracking error \eqref{tracking_error} converges to a small neighborhoods of zero, for all $ i= 1, \dots, n$. 

Note that the control objective can also be seen as a tracking control problem, so in the remainder of the paper the meaning of agents and followers is synonymous.

\section{Distributed Learning Consensus Protocol Design and Analysis}
In this section, we show how to design distributed learning leader-follower consensus protocols based on GPR, such that the tracking consensus problem is solved. Firstly, GPR is introduced  (Sec. \ref{sec_GPR}); then a distributed learning approach based on GPR is proposed in Sec. \ref{Section_DGP}; Sec. \ref{sec_ana} finally analyzes the stability of the system.

\subsection{Gaussian Process Regression }\label{sec_GPR}
The Gaussian process (GP) is a stochastic process that assigns a joint Gaussian distribution to any finite subset $ \left\{ \boldsymbol{x}_{1}, \boldsymbol{x}_{2},\dots, \boldsymbol{x}_{M} \right\} \subset \mathbb{X}$ in a continuous domain \cite{rasmussen2003gaussian}. A GP can be interpreted as a `distribution over functions' and written 
\begin{equation}
f\left( \boldsymbol{x} \right) \sim \mathcal{GP} \left( m \left( \boldsymbol{x} \right), k\left(  \boldsymbol{x},\boldsymbol{x}'\right) \right)
\end{equation}
where, $ m \left( \boldsymbol{x} \right):  \mathbb{X} \rightarrow \mathbb{R} $ is the prior mean function and $ k\left( \boldsymbol{x},\boldsymbol{x}'\right): \mathbb{X} \times \mathbb{X} \rightarrow \mathbb{R} $ is the covariance function. $m(\cdot)$ can be used to incorporate a prior model, $k(\cdot, \cdot)$ encodes abstract concepts such as smoothness, periodicity, etc. In this paper, the covariance function is typically chosen as the squared exponential kernel
\begin{equation} \label{k}
k \left( \boldsymbol{x}, \boldsymbol{x}' \right) = \sigma_r^2 \exp \left( -\frac{1}{2} \sum_{i}^n d_i \left( x_i - x_i' \right) \right),
\end{equation} 
where $\sigma_r^2 \in \mathbb{R}_{0,+}$, $d_i \in \mathbb{R}_+$. 

In order to illustrate the GPR, we assume a training data set $ \mathcal{D} = \left\{ \boldsymbol{X}, \boldsymbol{Y} \right\} $ consisting of training inputs $ \boldsymbol{X} = \left\{ \boldsymbol{x}^{\left(1\right)}, \boldsymbol{x}^{\left(2\right)}, \dots
\boldsymbol{x}^{\left(M\right)} \right\} \in \mathbb{R}^{m\times M} $ and training outputs $ \boldsymbol{Y} = \left\{ y^{\left(1\right)}, y^{\left(2\right)}, \dots, y^{\left(M\right)} \right\} \in \mathbb{R}^{M} $, which consists of noisy observations $ y^{\left(i\right)} = f\left( \boldsymbol{x}^{\left(i\right)} \right) + \boldsymbol{\varsigma} $, where $ i=1,\dots,M $, of an unknown function $f(\cdot)$ perturbed by  Gaussian noise, $ \boldsymbol{\varsigma} \sim \mathcal{N} \left( 0,  \sigma_o^2 \right) \in \mathbb{R}$. The evaluations of $y$ at a given test input $ \boldsymbol{x} \in \mathbb{R}^n $ is again a Gaussian distribution with the posterior mean and variance
\begin{align}
    \mu \left(\boldsymbol{x}, \mathcal{D}  \right) &= \boldsymbol{k}^\top\left( \boldsymbol{x} \right) \left( \boldsymbol{K}\left(\boldsymbol{X} \right) + \boldsymbol{I}_n \sigma_o^2 \right)^{-1} \boldsymbol{Y} ,\\ 
\sigma^2 \left(\boldsymbol{x}, \mathcal{D}  \right) &= k\left( \boldsymbol{x}, \boldsymbol{x} \right) \!- \!\boldsymbol{k}^\top \!\left( \boldsymbol{x} \right) \left( \boldsymbol{K}\left( \boldsymbol{X} \right) \!+ \!\boldsymbol{I}_n \sigma_o^2 \right)^{-1} \!\boldsymbol{k}\left( \boldsymbol{x} \right), 
\end{align}
where the elements of $\boldsymbol{k}\left ( \boldsymbol{x} \right )\in\mathbb{R}^{m}$ and $\boldsymbol{K} (\boldsymbol{X})\in\mathbb{R}^{m\times m}$ are defined through $\boldsymbol{k}_i(\boldsymbol{x})=k(\boldsymbol{x},\boldsymbol{X}^{(i)})$ and $K_{ii'}=k(\boldsymbol{X}^{(i)},\boldsymbol{X}^{(i')})$, respectively. 

In order to ensure a posterior variance function which captures the epistemic uncertainty properly, we assume a well-calibrated prior distribution:
\begin{assumption} [{\cite{lederer2019uniform} }] \label{ass_f_l}
Assume the nonlinear function $f\left ( \boldsymbol{x} \right )$ with Lipschitz constant $L_f$ to be a sample obtained from a Gaussian process $ \mathcal{GP}\left ( 0, k\left ( \boldsymbol{x},\boldsymbol{x}' \right )  \right )  $ with Lipschitz continuous kernel $k:\mathbb{R}^{m}\times\mathbb{R}^{m}\rightarrow\mathbb{R}_{0,+}$. 
\end{assumption}

\subsection{Distributed Learning based on GPR} \label{Section_DGP}
 Consider the nonlinear MAS in \eqref{followers} with unknown dynamics and unknown disturbance. The unknown function of the $i$-th agent to be predicted  based on GPR can be written as 
\begin{equation} \label{residual_dynamics}
    {\tau}\left ( x_i \right )  = f\left ( x_i \right ) + h \left ( x_i \right ), \quad i=1,2,\ldots,n.
\end{equation}

\begin{remark} \label{re_tau}
If prior knowledge on the dynamics of agents is available, (cf. Remark \ref{rem_residual_dynamics}) then, instead of \eqref{residual_dynamics}, the following unknown function
\begin{equation*}\label{residual_dynamics2}
    {\tau}\left ( x_i \right )  = f\left ( x_i \right ) - \hat{f}\left ( x_i \right ) + h \left ( x_i \right )
\end{equation*}
for $i= 1, \dots, n$ is considered, where $\hat{f}\left ( x_i  \right )$ represents the known part of the dynamics. Note that $\hat{f}\left ( \cdot  \right )$ does not need to be identical to the true dynamics ${f}\left ( \cdot  \right )$, but just a globally bounded term.
Accordingly, instead of the  protocol \eqref{controller}, (cf. Remark \ref{rem_residual_dynamics}), the  distributed protocol is then given by
\begin{equation*} \label{controller2}
u_i   = -k_i \xi_i  - \tilde{\mu}_i\left ( x_i \right )- \hat{f}\left ( x_i \right ), \quad i= 1, \dots, n.
\end{equation*}
\end{remark}

The straightforward approach to individually predict the unknown function \eqref{residual_dynamics}  is that each agent provides the posterior mean and variance of the associated GPs based on its own local data. However, in order to utilize the characteristics of information exchange in MAS, each agent's posterior knowledge is shared with its neighboring agents. This means that each agent not only exchanges state information, but also exchanges the prediction of its own GPs.

In \cite{deisenroth2015distributed}, the distributed Gaussian process (DGP) technique indeed exchanges the predictions of the agents. However, a central entity is required to distribute the final prediction. This means that the DGP approach is not a fully distributed learning approach. To overcome this drawback, we propose a fully distributed learning approach based on GPR. The prediction of each agent is computed by exchanging its own prediction with that of the neighboring agents. Considering the unknown function \eqref{residual_dynamics} of each agent,  GPs are trained with $M_i$ data pairs of set $\mathcal{D} _i  = \left\{ \bm{X}_i, \bm{Y}_i \right\}^{M_i} $, $i= 1,2,\dots , n$. The agent's posterior mean and variance of the multi-variable GP trained with data set $ \mathcal{D} _i $ are denoted as  $ \mu_i \left ( x,\mathcal{D}_i \right ) $ and $ \sigma_i^2 \left (  x,\mathcal{D}_i \right ) $, respectively. Thus the prediction of the $k$-th dimensional unknown function, $k=1,\dots, m$, is calculated as follows

\begin{equation}\label{mu_DGP}
   \begin{split}
     \tilde{\mu}_{ik}\left ( x_{i} \right ) \! &= \!\frac{ {\sigma}_{ik}^{-2}\left ( x_{i} \right ) \! {\mu}_{ik}\left ( x_{i} \right )  \! + \! \sum_{j =1}^{n} a_{ij} {\sigma}_{jk}^{-2}\left ( x_{i} \right ) \!  {\mu}_{jk}\left ( x_{i} \right ) }{\tilde{\sigma}_{ik}^{-2}\left ( x_{i} \right ) } \\
     & = \sum_{j=1}^{n} w_{{ik}}^{jk}\left ( x_{i} \right ) \mu_{jk} \left ( x_{i} \right )
   \end{split} 
\end{equation}
where $ \tilde{\sigma}_{ik}^{-2}\left ( x_{i} \right ) \!=\! {\sigma}_{ik}^{-2}\left ( x_{i}  \right ) \! +\! \sum_{j =1}^{n}  a_{ij} {\sigma}_{jk}^{-2}\left ( x_{i}  \right )$, $ {\mu}_{ik}\left ( x_{i}  \right )$ is $  \mu_{ik} \left ( \tau_k \mid x_i ,\mathcal{D}_i \right ) $ and $ {\sigma}_{ik}^{-2}\left ( x_{i}  \right )  $ is $ \sigma_{ik}^{-2} \left ( \tau_k \mid x_{i} ,\mathcal{D}_i \right )  $ for short, respectively. $ w_{ik}^{jk}\left ( x_{i}  \right ) = a_{ij} {{\sigma}_{jk}^{-2}\left ( x_{i}  \right )}{\tilde{\sigma}_{ik}^{2}\left ( x_{i}  \right )} \in \mathbb{R}_{0,+}$, for $ j =1, \dots,n $, and $w_{ik}^{ik}\left ( x_{i}  \right ) = {{\sigma}_{ik}^{-2}\left ( x_{i}  \right )}{\tilde{\sigma}_{ik}^{2}\left ( x_{i}  \right )} \in \mathbb{R}_{0,+}$ for $ i = 1, \dots,n $. Note that $\sum_{j=1}^n w_{ik}^{jk}\left ( x_{i}  \right ) =1$ for each agent $i$ and the elements of the weighted adjacency matrix $a_{ij}$ are governing the information exchange between neighbors.

According to the proposed distributed learning approach, we now show the bounded error of the prediction $\tilde{\mu}_{ik}$ associated with the $k$-th dimensional unknown function for each agent $i$. 

\begin{lemma} \label{lem_p}
For any compact set $  \Omega \in \mathbb{R}^m $ and a probability $ \delta \in \left( 0,1 \right) $, consider the unknown function \eqref{residual_dynamics} and GPs with the training data set $ \mathcal{D}_j = \left\{ \boldsymbol{X}_j, \boldsymbol{Y}_j \right\}^{M_j} $ containing $M_j$ data pairs satisfying \cref{ass_f_l}, $ \forall j = 1,2,\dots,n $. Moreover, consider the distributed learning with GP method in \eqref{mu_DGP}  to predict the $k$-th dimensional of \eqref{residual_dynamics}, $k=1,\dots,m$. Pick $\rho\in\mathbb{R}_+$ and define

\begin{align}
\label{eq:beta}
    \beta(\rho,\delta) &= 2m\log \left( \frac{r_{\Omega}\sqrt{m}}{2\rho} \right) + 2\log(n) - 2\log(\delta)\\
    \gamma_{jk}(\rho) &= (L_f+L_{\mu_{jk}})\rho+\sqrt{\beta(\rho,\delta)L_{\sigma_{jk}^2}\rho},
    \label{eq:gamma_i}
\end{align}
where $r_{\Omega}=\max_{{x},{x}'\in\Omega}\|{x}-{x}'\|$, $L_{\mu_{jk}}$ and $L_{\sigma_{jk}^2}$ are the 
Lipschitz constants of the individual GP mean and variance functions, respectively. Define the model error for the model estimate as $\Delta \tau(x_i) = \left \| \tau({x_i})- \tilde{\mu}({x_i}) \right \|$, and its $k$-th element $\Delta \tau(x_i)$ is denoted as $\Delta \tau_{k}(x_i)$. Then for agent $i$, there holds that
\begin{equation}
    \begin{split}
    \label{eq:un_bound}
    &P\Big \{ \Delta \tau_{k}(x_i) = \left | \tau_{k}({x_i})- \tilde{\mu}_{k}({x_i}) \right | \\
    & \leq \sum\limits_{j=1}^n w_{ik}^{jk} ({x_i})( \sqrt{ \beta(\rho,\delta)} \sigma_{jk}({x_i})\!+\!\gamma_{jk}(\rho) ), \forall x_i\in \mathbb{X} \Big \} \! \ge \! 1\!-\!\delta.
    \end{split}    
\end{equation}
\end{lemma}

By recalling \eqref{mu_DGP}, it is straightforward to show that
\begin{equation}
    \begin{split}
    |\tau_{ik}({x_i})-\tilde{\mu}_{ik}({x_i})| & = \left| \sum\limits_{j=1}^{n} w_{ik}^{jk}({x_i})\left (\mu_{jk} ({x_i}) - \tau_{ik}({x_i}) \right ) \right|\\
    & \leq \sum\limits_{j=1}^{n} w_{ik}^{jk}({x_i}) |\mu_{jk}({x_i})-\tau_{ik}({x_i})|,
    \end{split}
\end{equation}
where the equality in the first line is due to the fact that $\sum_{j=1}^n w_{ik}^{jk}\left ( x_i \right ) =1$, and the inequality in the second line follows from the triangle inequality. Under Assumption \ref{ass_f_l} we can apply \cite[Theorem 3.1]{lederer2019uniform} to the local models, such that we have with probability of at least $1-\delta/n$
\begin{align}
    |\tau_{k}({x_i})-\mu_{jk}({x_i})|\leq \sqrt{\beta(\rho,\delta)}\sigma_{jk}({x_i})+\gamma_{jk}(\rho)
\end{align}
for $\rho\in\mathbb{R}_+$ and $\beta(\rho,\delta)=2\log(M(\rho,\Omega)n/\delta)$, where $M(\rho,\Omega)$ denotes the $\rho$-covering number of $\Omega$. By overapproximating $\Omega$ through a hypercube with edge length $r_{\Omega}$, the covering number $M(\rho,\Omega)$ can be bounded by $(r_{\Omega}\sqrt{m}/(2\rho))^{m}$ satisfying \eqref{eq:beta}. Therefore, the joint probability over all agents yields the result in combination with the union bound.

\subsection{ Stability Analysis }\label{sec_ana}
In this subsection, we analyze the stability of MAS based on the proposed distributed protocol via Lyapunov analysis. Before giving the main result in Theorem \ref{theorem1}, we first present the following technical lemma.

\begin{lemma} \label{lem_V}
 If there exists a positive definite matrix $\tilde{L} = L+B $ satisfying Lemma \ref{lem_tidleL}, the smooth scalar function $V =  \boldsymbol{e}^\top \left [\tilde{L} \otimes I_m \right ] \boldsymbol{e}$ is bounded by
 \begin{equation} \label{Mbounded}
{\lambda_{\min} \left ( \tilde{L}^{-1} \right ) } \sum_{i=1}^{n} \left \| {\xi}_i \right \|^2 \leq V \leq  {\lambda_{\max} \left ( \tilde{L}^{-1} \right ) } \sum_{i=1}^{n} \left \| {\xi}_i \right \|^2,
\end{equation}
where $ \boldsymbol{e} = \left [ e_1^\top,e_2^\top,\dots, e_n^\top \right ]^\top $.
\end{lemma}

Since $\tilde{L}$ is positive definite, it holds that $\tilde{L}$ $\tilde{L}^{-1} = I_n$. Then, the smooth scalar function $V$ can be rewritten as
\begin{equation}
   \begin{split}
       V &=  \boldsymbol{e}^\top \left (\tilde{L} \otimes I_m \right ) \boldsymbol{e} = \boldsymbol{e}^\top \left (\tilde{L} \tilde{L}^{-1} \tilde{L} \otimes I_m \right ) \boldsymbol{e}
   \end{split} 
\end{equation}
Due to $ \boldsymbol{\xi}(k) =  \left ( \tilde{L} \otimes I_m \right ) \boldsymbol{e}(k) $, it holds that
\begin{equation}
   \begin{split}
       V = \boldsymbol{\xi}^\top \left( \tilde{L}^{-1}  \otimes I_m \right ) \boldsymbol{\xi}, 
   \end{split} 
\end{equation}
where $ \boldsymbol{\xi} = [\xi_1^\top, \dots, \xi_n^\top  ]^\top $.
So we have 
\[
\lambda_{\min} \left ( \tilde{L}^{-1} \right )  \sum_{i=1}^{n} \left \| \xi_i \right \|^2 \leq V \leq  \lambda_{\max} \left ( \tilde{L}^{-1} \right ) \sum_{i=1}^{n} \left \| \xi_i \right \|^2.
\]

Based on these lemmas, we are now in the position to state the main result of this paper.
\begin{theorem} \label{theorem1}
Consider a nonlinear MAS \eqref{followers} with a virtual leader \eqref{leader} under Assumptions \ref{bounded_fi}-\ref{ass_f_l}. Then, for any $k_i >0$, the distributed learning control law \eqref{controller} employing predictions \eqref{mu_DGP} based on the agent data sets $\mathcal{D}_i$, $i=1,\dots,n$,  achieves practical consensus, and, with  probability $(1-\delta)^m$, $\delta \in (0, 1)$, the tracking error \eqref{tracking_error} converges to a ball centered at the origin with the radius
\begin{equation} \label{radius}
   r = \frac{\sqrt{2\nu}}{ k^*\lambda_{\min } \left ( \tilde{L} \right )} ,
\end{equation}
where the parameters $k^*= \min \left \{ k_1, \dots, k_n\right \}$, $ \nu = \sum_{i=1}^{n} {\left (\Bar{f}_l^2 + \left \|\Delta \tau(x_i)\right \|^2\right )}$ and the model error $\Delta \tau(x_i)$ is defined in \cref{lem_p}.
\end{theorem}

A Lyapunov  candidate is chosen as
\begin{equation}
    V   = \frac{1}{2} \boldsymbol{e}^\top \left ( \tilde{L} \otimes I_m \right ) \boldsymbol{e},
\end{equation}
where the definition of $\tilde{L}$ is the same as Lemma \ref{lem_tidleL} and the global tracking error $\boldsymbol{e}$ is defined in Lemma \ref{lem_V}.

The time derivative of $V\left( t \right)$ along \eqref{e} is
\begin{equation}
    \begin{split}
      \dot{V}   &= \boldsymbol{e}^\top \left ( \tilde{L}\otimes I_m \right ) \dot{\boldsymbol{e}} = \sum_{i=1}^{n} \xi_i  ^\top  \dot{e}_i   \\
      &= \sum_{i=1}^{n} \xi_i  ^\top \Big( f \left ( x_i   \right ) + u_i   + h \left( x_i \right)     - {f}_l\left ( x_l  ,t\right ) \Big). 
    \end{split}
\end{equation}
Based on Assumption \ref{bounded_fl}, $\dot{V}  $ can be rewritten as
\begin{equation} \label{dotV2}
    \begin{split}
    \dot{V}   \leq \sum_{i=1}^{n} &  \Big\{ \xi_i   ^\top \Big(  f   \left ( x_i   \right )  +  u_i    + h \left ( x_i    \right ) \Big)  - \Bar{f}_l \left \| \xi_i    \right \|  \Big\}.
    \end{split}    
\end{equation}
Substituting the controller \eqref{controller} into \eqref{dotV2}, we have
\begin{equation} \label{dotV4}
  \begin{split}
    \dot{V}   &\leq \sum_{i=1}^{n} \left( -k_i \xi_i  ^\top \xi_i   + \xi_i  ^\top {\tau}_i \left( x_i   \right) - \xi_i  ^\top  \tilde{\mu}_i\left( x_i   \right)  - \Bar{f}_l \left \| \xi_i   \right \|  \right).
  \end{split} 
\end{equation}
Combining the scalar case prediction error bound proposed in Lemma \ref{lem_p} and the full component case prediction error bound proposed in \cite{umlauft2018uncertainty} (Lemma 2), due to the m-dimensional unknown function \eqref{residual_dynamics} it is straightforward to show that the overall upper bound for $\dot{V}$ with probability at least $ (1- \delta)^m $ holds that
\begin{equation} \label{dotV6}
  \begin{split}
\dot{V}      & \leq   - \sum_{i=1}^{n} k_i \left \| \xi_i   \right \|^2  + \sum_{i=1}^{n} \Big ( \left \lVert \Delta \tau(x_i) \right \rVert  -\Bar{f}_l \Big )   \left \| \xi_i   \right \|.
  \end{split} 
\end{equation}
Applying the inequality $v_1 \left \| x  \right \| \leq v_1^2/v_2 + v_2\left \| x \right \|^2/4 $ that holds $\forall x \in \mathbb{R}^m $ and $v_1, v_2 \in \mathbb{R}_+$, we have
\begin{equation} \label{Young2}
  \begin{split}
    \Big( \left \lVert \Delta \tau(x_i) \right \rVert  -\Bar{f}_l \Big )   \left \| \xi_i   \right \| \leq \frac{\left \|\Delta \tau(x_i)\right \|^2}{k_i} +\frac{\Bar{f}_l^2 }{k_i} + \frac{k_i}{2}\left \| \xi_i   \right \|^2.
  \end{split}     
\end{equation}
With \eqref{Young2}, equation \eqref{dotV6} can be rewritten as
\begin{equation} \label{dotV7}
  \dot{V}     \leq  -\frac{1}{2}\sum_{i=1}^{n}k_i  \left \| \xi_i   \right \|^2 +  \frac{\nu}{k^*}  ,
\end{equation}
with probability at least $ (1 \!-\! \delta)^m $, where $ \nu \!={\sum_{i=1}^{n}  {\left(\Bar{f}_l^2\!+\! \left \|\Delta \tau(x_i)\right \|^2 \right )}}  $, $k^* = \min \left \{ k_1, \dots, k_n\right \}$. According to \cref{lem_V} and \eqref{dotV7}, we have 
\[
P\left \{ \dot{V} \leq -  {k^*}{\lambda _{\min } \left ( \tilde{L} \right ) } \| e \|^2  +\frac{\nu}{k_*}  \right \} \ge(1- \delta)^m.
\]
 We have used the fact that $ 1/ \lambda _{\max}   ( \tilde{L}^{-1}  )=\lambda_{\min}   (\tilde{L})$. It then follows that, with  probability $(1-\delta)^m$, $\delta \in (0, 1)$,  the tracking error \eqref{tracking_error} converges to a ball centered at the origin  with radius:
\begin{equation*} 
   r = \frac{\sqrt{2\nu}}{ k^*\lambda_{\min } \left ( \tilde{L} \right )} .
\end{equation*}
This completes the proof.

\begin{remark}
By observing  equation \eqref{radius}, the error bound is related to the learning performance (characterized by $\nu$), the control gain (characterized by $k^*$) and the connectivity of the network (characterized by $\lambda_{\min}(\tilde{L})$). For control systems with given control gains,  we can reduce the tracking error by either improving the learning performance, or increasing the network connectivity.
\end{remark}

\begin{remark}
The result of Theorem 1 also holds analogously for the \emph{partial} dynamics knowledge case by applying the proposed controller \eqref{controller2} (cf. Remarks 1, 2 and 3).
\end{remark}

\section{Simulation}

In this subsection comparative simulations are presented, which contains three different control protocols. The first simulation is using a standard control protocol without GP learning, where the controller $ u_i   = -k_i \xi_i$; the second one with individual learning, which means each agent makes its own prediction of the GP independently with local training data only; the third one with proposed distributed learning based on GPR in this paper. We consider four agents with identical dynamics. To demonstrate the comparative simulations, we consider the following nonlinear agent dynamics 
\begin{equation} 
\begin{split}
     \dot{x}_{i1}    &= 2x_{i2}\sin \left ( x_{i1}   \right ) + u_{i1}, \\
     \dot{x}_{i2}    &= x_{i1}\cos \left (0.2 x_{i2}  ^2 + x_{i2}   \right ) + u_{i2},
\end{split}
\end{equation}
where $i= 1,\dots, 4$. 

The 400 training data pairs are equally distributed on the set $ [-2,2] \times [-2, 2] $. To make four different training sets for the four agents, we divide the training set directly into quarters for convenience. The initial positions of the four agents are chosen randomly within the interval $\left [ -2,2 \right ] $. The trajectory of the virtual leader is given as follows
\begin{equation} 
\begin{split}
\dot{x}_{l1}(t) = \sin(0.02\pi t), \\
 \dot{x}_{l2}(t) = \cos(0.02\pi t). 
\end{split}
\end{equation}
The environmental interference dynamics is chosen as
\begin{equation} 
\begin{split}
h_1 \left ( x_i\right ) =\sin \left ( x_{i2} \right ),\\
h_2 \left ( x_i \right )=\sin \left ( x_{i1} \right ).
\end{split}
\end{equation}

\begin{figure} [ht]
	\centering
	\includegraphics[width=1.6in]{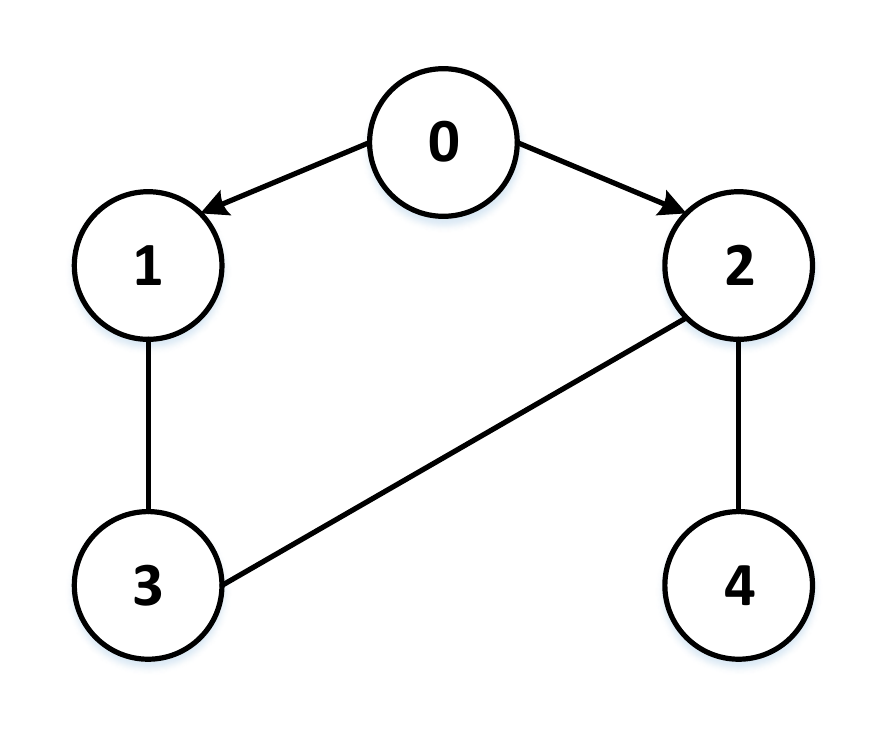}   	
	\caption{The communication graph of the MAS}
	\label{communication_topo}
\end{figure}

The control gains  are chosen to be $k_i = 2$. Fig. \ref{communication_topo} shows the connection relationship between the agents and the  virtual leader, which is chosen under Assumption \ref{ass_graph}. The diagonal matrix $B = {\rm diag} \left \{ 1,1,0,0,0 \right \} $, the adjacency matrix $A$ and matrix $\tilde{L}$ are given as follows
\[
A = \begin{bmatrix}
  0 & 0 & 1 & 0  \\
  0 & 0 & 1 & 1 \\
  1 & 1 & 0 & 0 \\
  0 & 1 & 0 & 0 
\end{bmatrix}, \ \tilde{L} = \begin{bmatrix}
  2 & 0 & -1 & 0  \\
  0 & 3 & -1 & -1 \\
  -1 & -1 & 2 & 0 \\
  0 & -1 & 0 & 1 
\end{bmatrix}.
\]

\begin{figure} [htp!]
	\centering
	\includegraphics[width=3.5in]{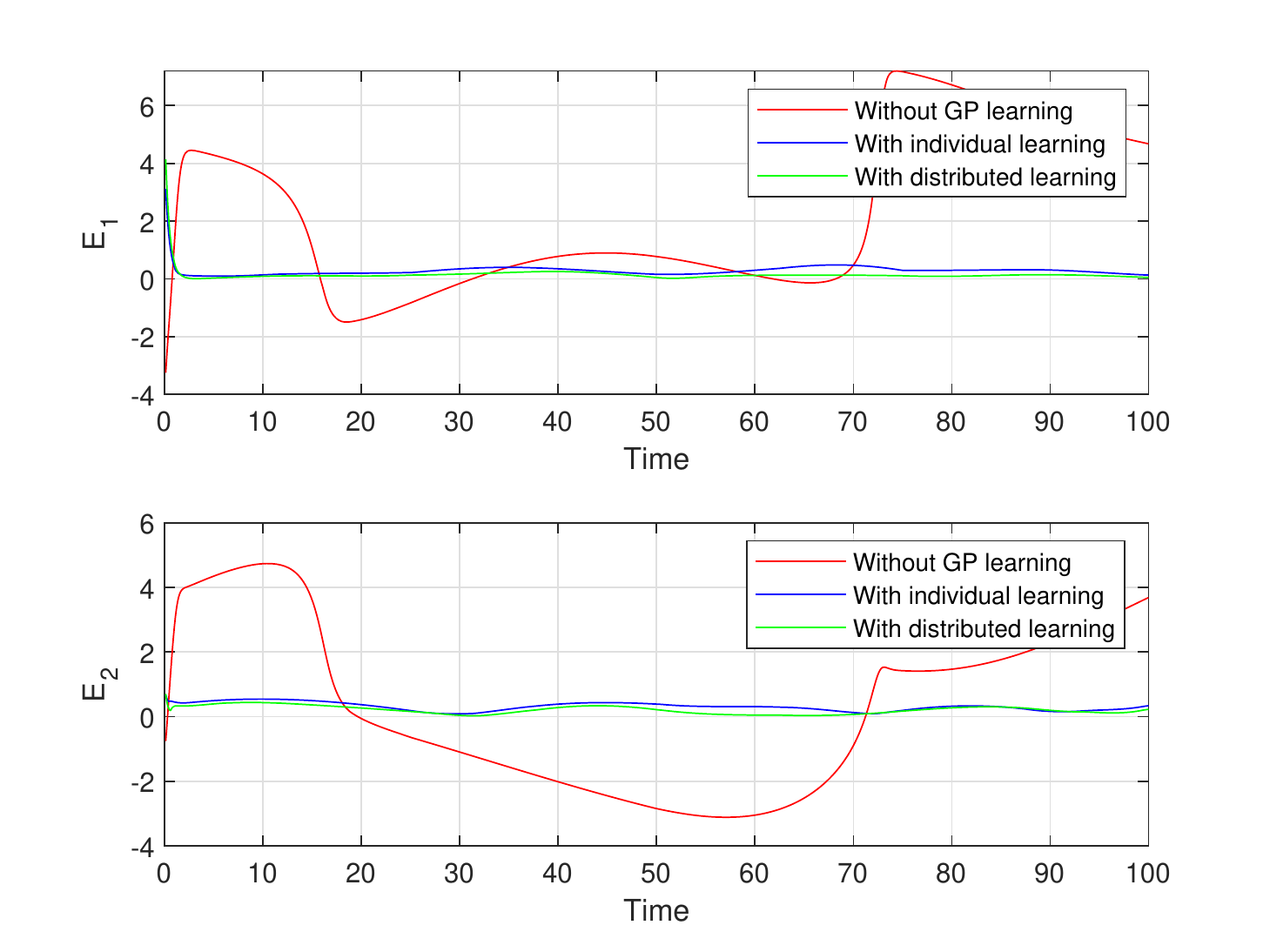}   	
	\caption{Accumulated errors of 4 agents curves for $x_{i1}$ and $x_{i2}$}
	\label{error}
\end{figure}

\begin{figure} [htp!]
	\centering
	\includegraphics[width=3.5in]{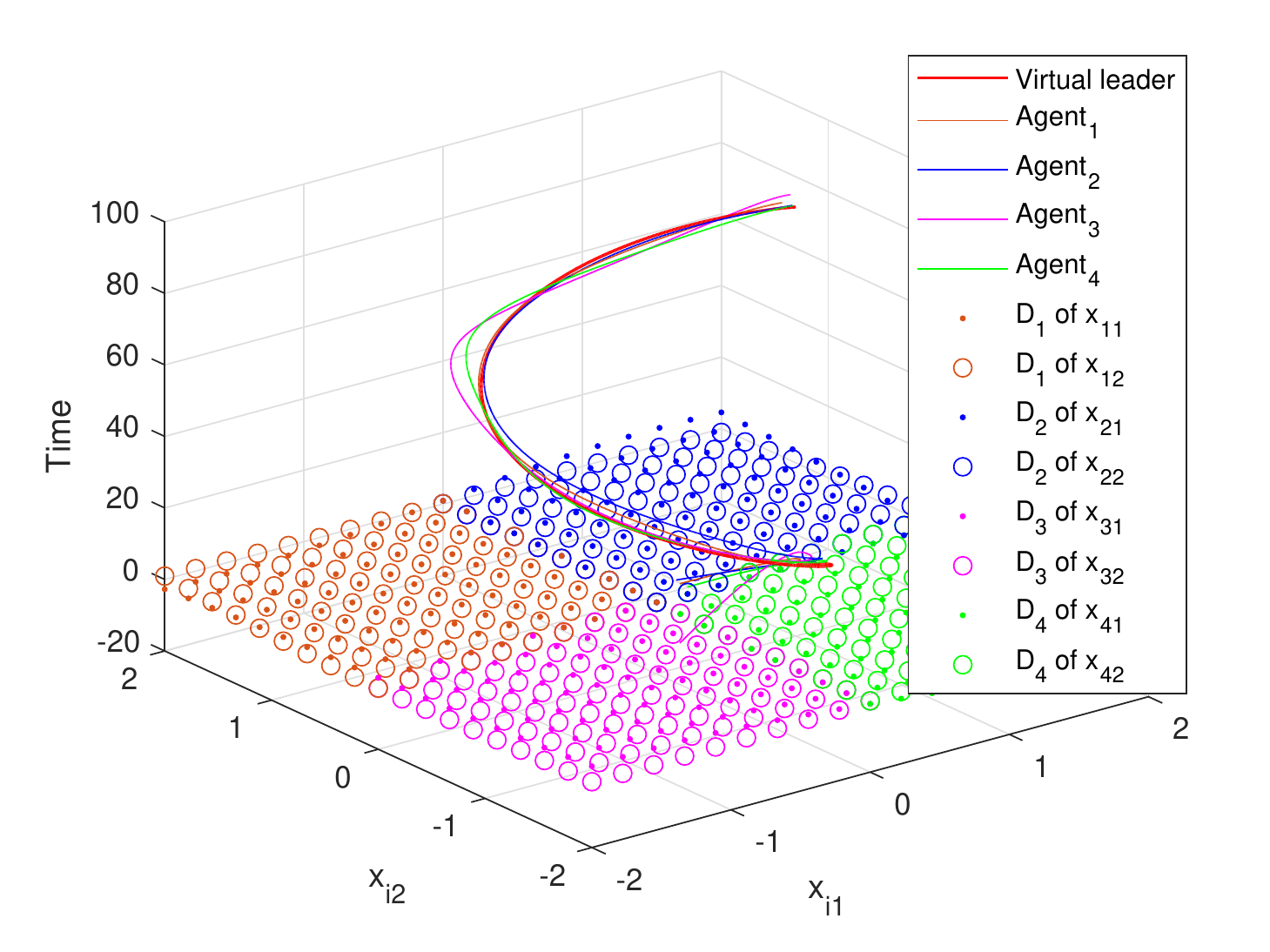}   	
	\caption{Three-dimensional trajectories of 4 agents with individual learning }
	\label{LGP}
\end{figure}

\begin{figure} [htp!]
	\centering
	\includegraphics[width=3.5in]{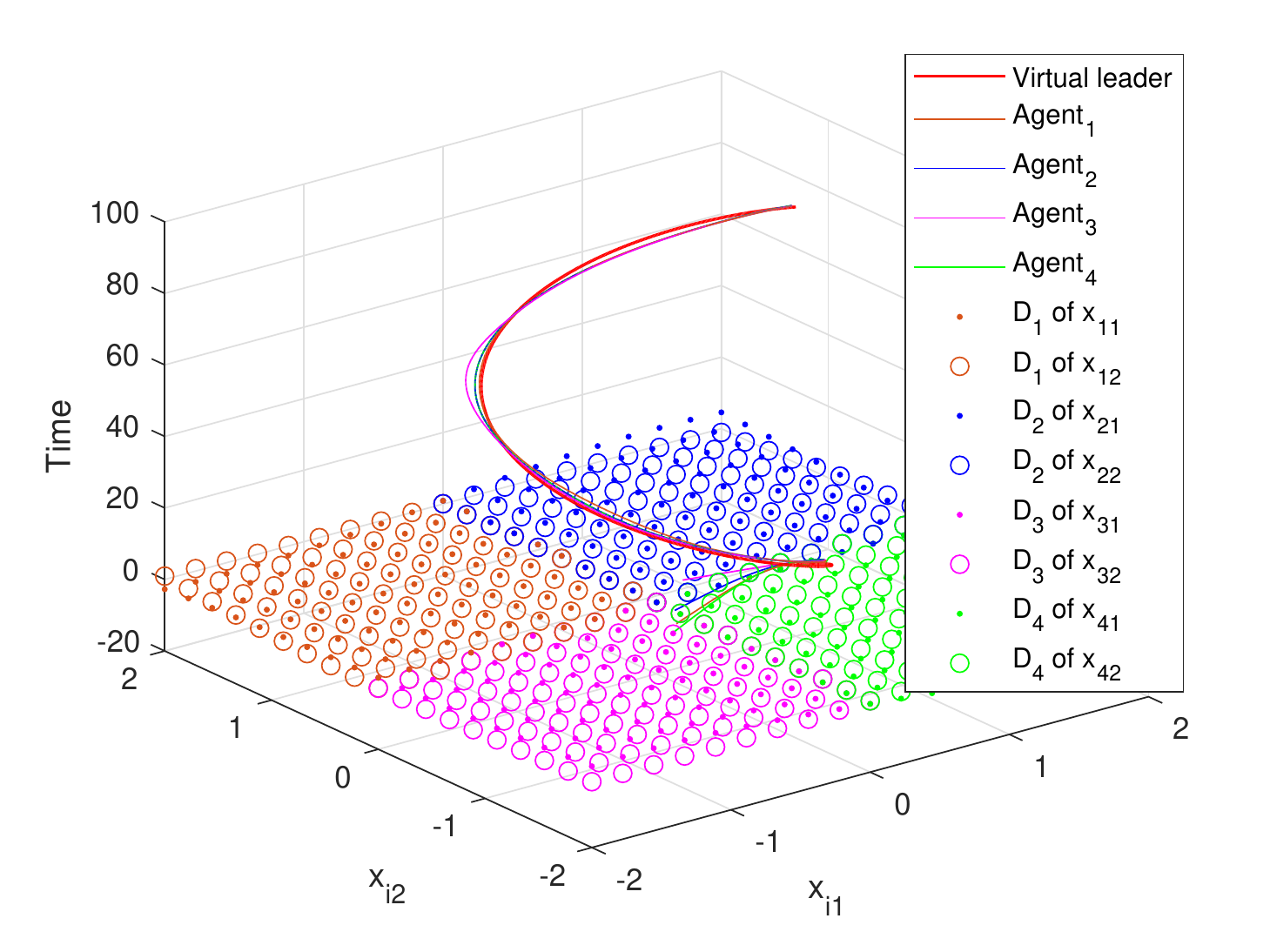}   	
	\caption{Three-dimensional trajectories of 4 agents with distributed learning based on GPR }
	\label{DGP}
\end{figure}

 The examples show the tracking performance in three cases, which include the case without learning of the unknown dunamics, with individual learning and with distributed learning based on GPR. \cref{error} shows the accumulated tracking errors of the MAS, which are defined as $E_j =  {\textstyle \sum_{i=1}^{4}} \left | {x}_{i,j} - {x}_l \right | $, where $j=1,2$ denotes the dimension of $x$, and $i=1,\dots,4$ denotes the number of the agents. The tracking errors with the GP learning approaches are smaller than without learning approach. The curves of the accumulated errors  with distributed learning tend to zero and have the smallest amplitudes, which show that MAS with distributed learning achieve the best tracking performance. To visually demonstrate the tracking performance, the three-dimensional plots of the motion curves of agents, which represent the evolution of the states over time, are shown in \cref{LGP} and \cref{DGP}, respectively. All four agents can follow the virtual leader, however the tracking trajectories do not converge very well all the time in \cref{LGP} compared to the approach with distributed learning in Fig. \ref{DGP}.

\section{Conclusions}

In this paper, we have proposed a consensus control protocol with distributed learning based on GPR for an unknown nonlinear MAS. We have first provided a distributed learning approach based on GPR to estimate the unknown agent dynamics. Unlike the individual learning approach, each agent exchanges the posterior knowledge of its own GPs, which overcomes the drawback of the centralized learning-based approach. Based on the estimated models, we then have designed a distributed protocol which guarantees that the states of the agents track that of the leader and the tracking error of the controlled MAS to converge to a ball centered at origin with a high probability. The radius of the error ball depends on the control gains, the smallest eigenvalue of a matrix that involving the communication graph of MAS and the learning performance.






\bibliography{M012_ref} 

\begin{thebibliography}{10}
\providecommand{\url}[1]{#1}
\csname url@samestyle\endcsname
\providecommand{\newblock}{\relax}
\providecommand{\bibinfo}[2]{#2}
\providecommand{\BIBentrySTDinterwordspacing}{\spaceskip=0pt\relax}
\providecommand{\BIBentryALTinterwordstretchfactor}{4}
\providecommand{\BIBentryALTinterwordspacing}{\spaceskip=\fontdimen2\font plus
\BIBentryALTinterwordstretchfactor\fontdimen3\font minus
  \fontdimen4\font\relax}
\providecommand{\BIBforeignlanguage}[2]{{%
\expandafter\ifx\csname l@#1\endcsname\relax
\typeout{** WARNING: IEEEtran.bst: No hyphenation pattern has been}%
\typeout{** loaded for the language `#1'. Using the pattern for}%
\typeout{** the default language instead.}%
\else
\language=\csname l@#1\endcsname
\fi
#2}}
\providecommand{\BIBdecl}{\relax}
\BIBdecl

\bibitem{oriolo2002wmr}
G.~Oriolo, A.~De~Luca, and M.~Vendittelli, ``{WMR} control via dynamic feedback
  linearization: design, implementation, and experimental validation,''
  \emph{IEEE Transactions on Control Systems Technology}, vol.~10, no.~6, pp.
  835--852, 2002.

\bibitem{van2018adaptive}
M.~Van, M.~Mavrovouniotis, and S.~S. Ge, ``An adaptive backstepping nonsingular
  fast terminal sliding mode control for robust fault tolerant control of robot
  manipulators,'' \emph{IEEE Transactions on Systems, Man, and Cybernetics:
  Systems}, vol.~49, no.~7, pp. 1448--1458, 2018.

\bibitem{shen2017trajectory}
C.~Shen, Y.~Shi, and B.~Buckham, ``Trajectory tracking control of an autonomous
  underwater vehicle using {L}yapunov-based model predictive control,''
  \emph{IEEE Transactions on Industrial Electronics}, vol.~65, no.~7, pp.
  5796--5805, 2017.

\bibitem{umlauft2017learning}
J.~Umlauft, A.~Lederer, and S.~Hirche, ``Learning stable {G}aussian process
  state space models,'' in \emph{2017 American Control Conference (ACC)}.\hskip
  1em plus 0.5em minus 0.4em\relax IEEE, 2017, pp. 1499--1504.

\bibitem{lederer2019uniform}
A.~Lederer, J.~Umlauft, and S.~Hirche, ``Uniform error bounds for {G}aussian
  process regression with application to safe control,'' in \emph{Advances in
  Neural Information Processing Systems}, 2019, pp. 659--669.

\bibitem{umlauft2019feedback}
J.~Umlauft and S.~Hirche, ``Feedback linearization based on {G}aussian
  processes with event-triggered online learning,'' \emph{IEEE Transactions on
  Automatic Control}, vol.~65, no.~10, pp. 4154--4169, 2019.

\bibitem{cao2012overview}
Y.~Cao, W.~Yu, W.~Ren, and G.~Chen, ``An overview of recent progress in the
  study of distributed multi-agent coordination,'' \emph{IEEE Transactions on
  Industrial informatics}, vol.~9, no.~1, pp. 427--438, 2012.

\bibitem{oh2015survey}
K.-K. Oh, M.-C. Park, and H.-S. Ahn, ``A survey of multi-agent formation
  control,'' \emph{Automatica}, vol.~53, pp. 424--440, 2015.

\bibitem{abdulghafor2018overview}
R.~Abdulghafor, S.~S. Abdullah, S.~Turaev, and M.~Othman, ``An overview of the
  consensus problem in the control of multi-agent systems,'' \emph{Automatika},
  vol.~59, no.~2, pp. 143--157, 2018.

\bibitem{qian2018output}
Y.-Y. Qian, L.~Liu, and G.~Feng, ``Output consensus of heterogeneous linear
  multi-agent systems with adaptive event-triggered control,'' \emph{IEEE
  Transactions on Automatic Control}, vol.~64, no.~6, pp. 2606--2613, 2018.

\bibitem{li2019multilayer}
D.~Li, S.~S. Ge, W.~He, G.~Ma, and L.~Xie, ``Multilayer formation control of
  multi-agent systems,'' \emph{Automatica}, vol. 109, p. 108558, 2019.

\bibitem{chen2014adaptive}
C.~P. Chen, G.-X. Wen, Y.-J. Liu, and F.-Y. Wang, ``Adaptive consensus control
  for a class of nonlinear multiagent time-delay systems using neural
  networks,'' \emph{IEEE Transactions on Neural Networks and Learning Systems},
  vol.~25, no.~6, pp. 1217--1226, 2014.

\bibitem{el2014neuro}
S.~El-Ferik, A.~Qureshi, and F.~L. Lewis, ``Neuro-adaptive cooperative tracking
  control of unknown higher-order affine nonlinear systems,''
  \emph{Automatica}, vol.~50, no.~3, pp. 798--808, 2014.

\bibitem{zou2020finite}
W.~Zou, P.~Shi, Z.~Xiang, and Y.~Shi, ``Finite-time consensus of second-order
  switched nonlinear multi-agent systems,'' \emph{IEEE Transactions on Neural
  Networks and Learning Systems}, vol.~31, no.~5, pp. 1757--1762, 2020.

\bibitem{ouyang2018gaussian}
R.~Ouyang and K.~H. Low, ``Gaussian process decentralized data fusion meets
  transfer learning in large-scale distributed cooperative perception,'' in
  \emph{Proceedings of the AAAI Conference on Artificial Intelligence},
  vol.~32, no.~1, 2018.

\bibitem{nguyen2019exact}
D.-T. Nguyen, M.~Filippone, and P.~Michiardi, ``Exact {G}aussian process
  regression with distributed computations,'' in \emph{Proceedings of the 34th
  ACM/SIGAPP Symposium on Applied Computing}, 2019, pp. 1286--1295.

\bibitem{hoang2019collective}
T.~N. Hoang, Q.~M. Hoang, K.~H. Low, and J.~How, ``Collective online learning
  of gaussian processes in massive multi-agent systems,'' in \emph{Proceedings
  of the AAAI Conference on Artificial Intelligence}, vol.~33, no.~01, 2019,
  pp. 7850--7857.

\bibitem{yuan2020communication}
Z.~Yuan and M.~Zhu, ``Communication-aware distributed {G}aussian process
  regression algorithms for real-time machine learning,'' in \emph{2020
  American Control Conference (ACC)}.\hskip 1em plus 0.5em minus 0.4em\relax
  IEEE, 2020, pp. 2197--2202.

\bibitem{le2020gaussian}
V.-A. Le and T.~X. Nghiem, ``{G}aussian process based distributed model
  predictive control for multi-agent systems using sequential convex
  programming and {ADMM},'' in \emph{2020 IEEE Conference on Control Technology
  and Applications (CCTA)}.\hskip 1em plus 0.5em minus 0.4em\relax IEEE, 2020,
  pp. 31--36.

\bibitem{hong2006tracking}
Y.~Hong, J.~Hu, and L.~Gao, ``Tracking control for multi-agent consensus with
  an active leader and variable topology,'' \emph{Automatica}, vol.~42, no.~7,
  pp. 1177--1182, 2006.

\bibitem{rasmussen2003gaussian}
C.~E. Rasmussen and C.~K.~I. Williams, \emph{{G}aussian {P}rocesses for
  {M}achine {L}earning}.\hskip 1em plus 0.5em minus 0.4em\relax Cambridge, MA,
  USA: MIT Press, 2006.

\bibitem{deisenroth2015distributed}
M.~Deisenroth and J.~W. Ng, ``Distributed {G}aussian processes,'' in
  \emph{International Conference on Machine Learning}.\hskip 1em plus 0.5em
  minus 0.4em\relax PMLR, 2015, pp. 1481--1490.

\bibitem{umlauft2018uncertainty}
J.~Umlauft, L.~P{\"o}hler, and S.~Hirche, ``An uncertainty-based control
  {L}yapunov approach for control-affine systems modeled by {G}aussian
  process,'' \emph{IEEE Control Systems Letters}, vol.~2, no.~3, pp. 483--488,
  2018.

\end{thebibliography}

\end{document}